\documentclass[journal]{IEEEtran}

\ifCLASSINFOpdf
\else
\fi

\usepackage{comment}
\usepackage{algorithmic} 
\usepackage[ruled,vlined]{algorithm2e}

\usepackage{graphicx}
\usepackage{mathtools}
\usepackage{color}
\usepackage{placeins}
\usepackage{float}
\usepackage{hyperref}
\usepackage{tabularx,colortbl}

\usepackage{siunitx}
\usepackage{booktabs}
\usepackage{amssymb}
\usepackage{cite}

\usepackage{cuted, ragged2e}
\begin{document}
%
\title{Source detection via multi-label classification }
%
%
\author{Jayakrishnan Vijayamohanan,
        Arjun Gupta,
        Oameed Noakoasteen,
        Sotirios Goudos
        and~Christos Christodoulou
\thanks{This work has been submitted to the IEEE-OJSP for possible publication. Copyright may be transferred without notice, after which this version may no longer be accessible. Jayakrishnan Vijayamohanan (jayakrishnan@unm.edu), Arjun Gupta, Oameed Noakoasteen, and Christos Christodoulou are with the Department   
of Electrical and Computer Engineering, University of New Mexico, Albuquerque,
NM, 87131 USA. Sotirios Goudos is with Department of Physics, Aristotle University of Thessaloniki, 54124 Greece }
}


\maketitle

\begin{abstract}
     Radio source detection through conventional algorithms has been unreliable when trying to solve for large number of sources in the presence of low SINR and less number of snapshots. We address this by reformulating source detection as a multi-class classification problem solved using deep learning frameworks. Incoming waveforms are sampled using a centro-symmetric linear array with omni-directional elements and the normalized upper triangle of the autocorrelation matrix is extracted as the input feature to a modified convolutional neural network with uni-dimensional filters, trained to detect the sources in the presence of both uncorrelated and correlated signals. Two detection algorithms are introduced and referred to as CNNDetector and RadioNet, and subsequently benchmarked against the conventional source detection algorithms. By including pre-processing in forward backward spatial smoothing, RadioNet can also resolve the number of uncorrelated sources in the presence of correlated paths. Finally, the algorithms are stress tested under challenging operational conditions and extensive evaluations are presented showing the efficacy and contributions of the introduced predictive models.

\vspace{5mm}
\end{abstract}

\begin{IEEEkeywords}
Array Signal Processing, Multi-label Classification, ResNet, CNN, Direction of Arrival,Residual Learning, Signal Source Detection, Statistical Signal Processing
\end{IEEEkeywords}

\IEEEpeerreviewmaketitle

\section{Introduction}
\IEEEPARstart{R}{adio} events surround the world around us. We live in an increasingly connected environment fueled by hand held devices, household appliances, autonomous vehicles, and wearable devices; the ease of connectivity provided by multi-input multi-output (MIMO) systems and fast data rates of modern communication protocols. As a result, radio source detection has received significant attention over the years \cite{manel_book,van_trees,arjun-awpl,doa_liu,doa_imperfect}, and made its way into applications across various domains of science and engineering  \cite{doa,van_trees,massa_localization,robot_localization,arjun-eucap,ojsp}. High resolution direction of arrival (DoA) estimation algorithms such as MuSiC, root-MuSiC, ESPRIT and several others including non-parametric machine learning and deep learning methods require the knowledge of the number of sources to compute a viable localization estimation\cite{manel_book,van_trees,massa_doa,music,rootmusic,arjun-aps2020}. Maximum likelihood estimation (MLE) using a Gaussian assumption is considered the optimal solution to estimate the number of sources \cite{van_trees,fisher_mle,fisher_mle2}. However, MLE is computationally expensive and is not a viable solution for real-time source detection. The conventional approach for estimating the number of sources is to apply information theoretic criterion like the minimum description length (MDL) or Akaike information criterion (AIC) \cite{det-info_crit, det-info_crit2, det-info_crit3}. However, these approaches suffer degradation in detection performance in the presence of low Signal to Interference plus Noise Ratio (SINR), fewer number of samples, and an increasing number of sources to resolve \cite{detect_eigen}. Eigen methods involving the Eigen value decomposition (EVD) of the autocorrelation matrix have also been employed to estimate the number of sources with good success \cite{80959}.

More recently, deep neural network (DNN) architectures such as auto-encoder and long short-term memory (LSTM) have been shown to perform source localization with good accuracy \cite{doa_liu,manel_book,massa_review,doa_autoencoder, arjun-aps2020}. DNN based approaches have also been implemented to estimate source detection from the sampled waveform \cite{vanilla1,vanilla2,cnn_sourcedetect,feifeinet,tsp}. Some of the early works include vanilla networks capable of detecting the presence of no more than $4$ source signals \cite{vanilla1, vanilla2}. In \cite{cnn_sourcedetect}, a convolutional neural network (CNN) was trained to predict the number of sources by feeding it raw signal waveforms. This approach ignores the possibility of the autocorrelation being rank deficient, directly affecting the detection performance \cite{pillai}. This drawback was resolved in \cite{feifeinet}, where a neural network was trained by feeding the Eigen vectors obtained from the autocorrelation as the input. This approach improved generalization and the overall detection performance of the algorithm. This model was also able to resolve correlated sources by performing forward backward spatial smoothing (FBSS) on the autocorrelation matrix \cite{pillai}. However, computing the EVD can add additional computational complexity which can be avoided \cite{manel_book, arjun-awpl}. 

In this paper, we introduce CNN based frameworks to detect the number of sources without the need to compute EVD. Furthermore, we introduce residual layers to improve on the performance over vanilla CNNs. Instead of computing the EVD, we extract the upper triangular elements of the autocorrelation matrix as the input feature. Note that the autocorrelation is a symmetric matrix and this reduction preserves all the spatial-temporal cues obtained from the waveform \cite{christos_book, manel_book}. We proceed to stress test the methods and frameworks in the presence of both uncorrelated and correlated sources to find the performance threshold for the algorithms. The major contributions of this paper are summarized below:\\
1. We introduce a novel detection framework and evaluate the models up to the operational threshold of $L- 1$ sources present in the sampled waveform, with $L$ being the number of elements in the array. Literature pertaining to solving this problem considers only the presence of less than or equal to four or five sources to resolve \cite{vanilla1,vanilla2,cnn_sourcedetect,feifeinet,tsp}. \\
2. Comprehensive detection analysis is done by studying the sensitivity of the algorithms to a varying number of correlated and non-correlated sources. Section IV goes over experiments investigating the interactions between the number of correlated and uncorrelated sources and their effects on the detection probability.
 
The rest of the paper is organized in four parts. Section II describes the problem formulation in terms of the signal model and feature extraction. Section III introduces the CNNDetector and its strength and weakness are investigated to discover the shortcomings of the CNNDetector. In Section IV, we introduce residual learning in the form of RadioNet to improve on CNNDetector. We demonstrate the improvement provided by RadioNet and show the improvement in generalization provided over existing models and methods. Finally, the results are concluded in Section V.  

\section{Problem Formulation} \label{Data Model}

\subsection{Signal Model}

We begin the investigation around the ideal case of source detection in the absence of correlated or coherent sources and then extend to the more complex scenario of source detection in the presence of coherent sources. We consider a centro-symmetric linear array of $L$ identical antennas placed along an axis with the center of the array coinciding with the origin of the co-ordinate system. The aperture of this array is illuminated by $M$ non-coherent signals, $N$ number of snapshots are captured, and the source location $\theta_{M_i}$ is chosen arbitrarily from $-60^{o}$ $\leq$ $\theta_{M_i}$ $\leq 60^{o}$. The inter elemental spacing within the array is denoted by $\tilde{d_{i}}$. The complex envelope model of the received signal ${\bf X}(t)$ $\in$ $\mathbb{C}^{L\times N}$ can be modeled as,

\begin{equation}\label{eq:signalmodel}
{\bf X}(t)={\bf A}(\theta)~{\bf S}(t)+{\bf n}(t)
\end{equation}
where ${\bf A(\theta)} = [{\bf a}(\theta_1), {\bf a}(\theta_2), \cdots {\bf a}(\theta_{M})]$ $\in$ $\mathbb{C}^{L\times M}$ is the array manifold matrix, which contains steering vectors of the form
\begin{equation}\label{eq:steering_vector}
{\bf a}(\theta_m)=[ 1, e^{j2\pi \frac{d_1}{\lambda}sin(\theta_2)}, \cdots, e^{j 2 \pi\frac{d_{L-1}}{\lambda}sin(\theta_M)}]^T
\end{equation}

${\bf S(t)} \in  \mathbb{C}^{M \times N}$ consists of the independent signal vectors which can be denoted by $\bf s_1,...,\bf s_M$. These independent signals are generated by applying Quadrature Phase Shift Keying (QPSK) modulation to a random bit sequence.  ${\bf n}(t) \in \mathbb{C}^{L \times 1}$ is an additive white Gaussian noise (AWGN) vector \cite{arjun-aps2019}. $\lambda$ is the signal wavelength. 

This assumption of uncorrelated sources however is often too simplistic when modelling RF environments in dense urban areas. To accurately model such environments we need to introduce a varying number of correlated or coherent sources in the mix. If one of the original transmitted signal can be denoted by $s_1(t)$, then the $k$th coherent signal of $s_1(t)$ can be written as, 
\begin{equation}\label{eq:correlated_sig}
    {\bf{s}}_k(t) = \rho_ke^{j\phi_k}{\bf{s}}_1(t)
\end{equation}
where $\rho_k$ is the amplitude fading factor and $\phi_k$ is the phase change caused due to multi-path fading. The received signal ${\bf{S}}(t)$ can be modelled as a collection of both independent and correlated signals closely replicating the modern communication environment. ${\bf S}(t)$ $\in$ $\mathbb{C}^{M \times N}$ contains both the zero mean independent signals and the correlated signals generated using \ref{eq:correlated_sig}. 
The presence of correlated signals would result in the matrix being rank-deficient \cite{pillai}. To avoid this, we use FBSS to smooth the autocorrelation before extracting the feature vectors. To compute FBSS, the array is divided into $K$ overlapped subarrays with $K = L - L_0 + 1$, $L_0$ being the dimension of each sub-array. The smoothed autocorrelation matrix is given by, 
\begin{equation}
    {\bf{R_{fb}}} = \frac{1}{2K} \sum_{k=1}^K({\bf{R_{ff}}}^{(k)} + {\bf{R_{bb}}}^{(k)})
\end{equation}
where, ${\bf{R_{ff}}}^{(k)}$ and ${\bf{R_{bb}}}^{(k)}$ are the forward and backward autocorrelation matrices constructed from the $k$th subarray and ${\bf{R_{fb}}}$ is the spatially smoothed autocorrelation matrix.

\subsection{Feature Extraction}
Due to the symmetry of the autocorrelation matrix, extracting the upper triangular elements along with the diagonals should provide sufficient information for learning algorithms \cite{manel_book}. For each sensor, the real and imaginary part is separated, and then normalized to obtain a column vector of dimension, $(L * (L+1))$. This way the upper-triangle of the computed ${\bf{R}}_{xx}$, or ${\bf{R_{fb}}}$ in the case of correlated signals, becomes the input feature for the learning algorithms. Since we train the model using synthetic data generated by realistic simulations, we control the number of signals in sampled waveform for a given time instance which becomes the ground truth or the target variable \cite{manel_book}.

\section{Detection Framework I} 
We begin with the premise that the information on the number of sources is contained within the autocorrelation matrix obtained from the sampled waveform \cite{manel_book, van_trees}. We make use of the observation that a well designed neural network with substantial depth should be able to learn the mapping from the information present in the autocorrelation to the number of sources present in the sampled waveform as demonstrated in \cite{el_zhoogby,radionet,dl_coherent}.
\begin{figure*}
\center
\includegraphics[scale=0.6]{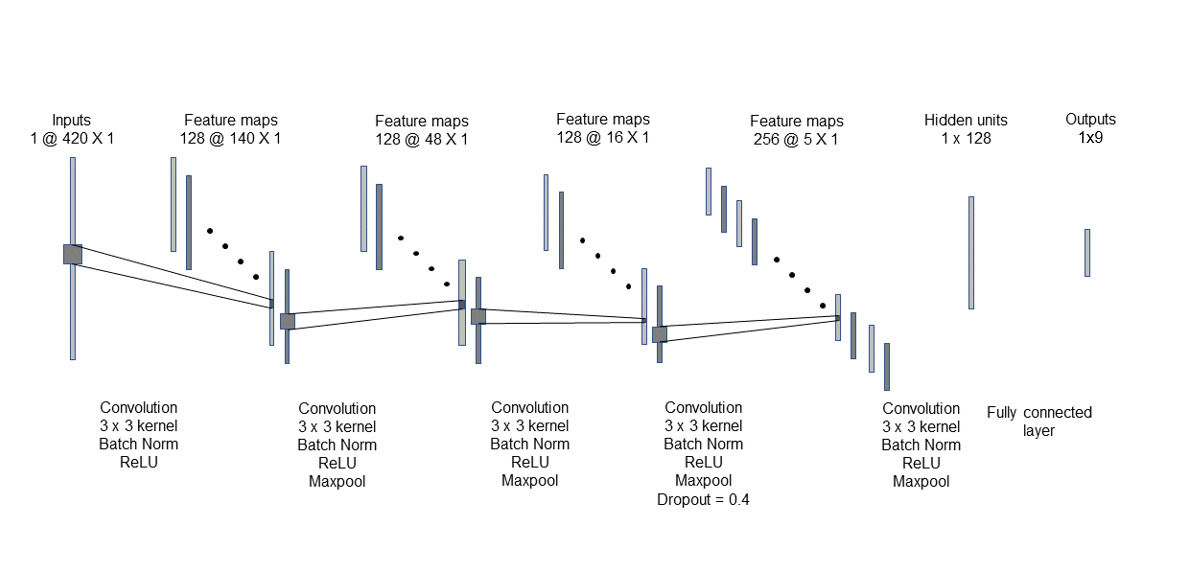}
\caption{Network architecture of the CNNDetector.}
\label{fig:CNNarch}
\end{figure*}

In this section we propose a detection framework developed with convolutional neural nets (CNN) as the core learning unit. CNNs are a class of neural network architecture which make use of filters to extract underlying information within the data and are capable of extracting highly abstract features from structured data such as images \cite{Goodfellow,manel_book, alexnet,resnet}. As such, they have been adopted and reformulated to perform various computer vision objectives such as object detection, localization and semantic segmentation. We start the detection architecture design with a rather simple stacked CNN framework referred to as the CNNDetector. 

\subsection{CNNDetector}

The CNNDetector architecture is composed of stacks of convolutional layers, followed by a fully connected layer to generate the discrete outputs in the form of a one-hot encoded target vector. To account for the uni-dimensional feature vector obtained from  feature extraction we employ 1D convolutions with uni-dimensional filters as visualized in Figure \ref{fig:CNNarch}. The kernel size is kept constant at $[1 \times 3]$, and batch normalization is performed after each layer to normalize the contribution to a layer for every mini-batch \cite{batch_normalization}. Batch normalization achieves fixed distributions of inputs and  prohibits internal covariate shift. Pooling is done in the last four convolutional layers. We optimize the number of layers along with the number of filters used in each layer. We found the $5$ stacked CNN layers provides enough depth for detection without overfitting on the data. The number of filter size was $128$, except in the fourth layer where it is increased to $256$. The fourth convolutional layer also includes a dropout rate of $40\%$ to facilitate generalization. The output layer was ket a constant at $[1 \times L -1]$, $L$ being the number of elements in the array. We minimize a categorical cross-entropy loss defined as,
\begin{equation}
\mathcal{L} = -\sum_{i = 0}^{L-1}y_{i} . log(\tilde{y}_i)
\end{equation}
here $\tilde{y}_i$  is the $i^{th}$ scalar value in the model output and $y_i$ is the corresponding target value \cite{cross_entropy}. The categorical cross-entropy is a measure of the difference between the discrete distributions corresponding to each possible class in the problem. Minimizing the negative loss ensures that the loss becomes increasingly smaller as the distributions converge. The network parameters were initialized using Xavier initialization. This is done so that the variance remains the same across every layer \cite{pmlr-v9-glorot10a}. Once the network is trained it is evaluated using the detection algorithm for uncorrelated sources presented in Algorithm \ref{algo:Source Detection using Detection Framework-I}.

\begin{algorithm}
  \begin{algorithmic}[1]\label{algo:Source Detection using Detection Framework-I}
    \STATE \textbf{Require: }{$\bf{X}$, $L$, $CNNDetector$ }\\
    \STATE \textbf{Outputs:  }{$y_{label}$}\\
    \STATE $temp$ $\in$$\mathbb{C}^{1 \times (L.(L+1/2))} \gets $ random temporary array
    \STATE k,l = 1\\ 
    \STATE $\bf{R_{xx}} = \bf{X}.\bf{X}^H$
    \FOR{$i \in\{1,\dots,L$\}}
        \FOR{$j \in \{l,\dots,L$\}}
            \STATE temp[k]  = $\bf{R_{xx}}$[i,j]
            \STATE k += 1
        \ENDFOR
        \STATE l += 1
    \ENDFOR\\
    \STATE i = 1 \\
    \FOR{$j$ $\in$ \{1,$\dots$,(L$\times$(L+1)/2)\}}
        \STATE feature vector[i] = $\mathbb{R}(temp[j])$\\
        \STATE i += 1\\
        \STATE feature vector[i] = $\mathbb{I}(temp[j])$\\
        \STATE i += 1\\
        \ENDFOR
   \STATE $y_{label}$ $\xleftarrow{}$ $CNN$ $detector$(feature vector)\\
   \RETURN $y_{label}$
  \end{algorithmic}
  \caption{Source number detection using CNNDetector}
\end{algorithm}

\subsection{Experiments and Evaluation}
We evaluate the framework in the context of source detection in the presence of uncorrleated sources only. Since we extracted the features from the autocorrelation matrix, the theoretical maximum number of sources this approach can resolve is capped at $L - 1$. We use a $10$ element array for the studies in this paper without any loss of generalization, the theories and methods can be readily extrapolated to arrays of any given size and shape. The data for training and testing the networks were generated using the equations introduced in Section \ref{Data Model}. $110,000$ frames were generated and further partitioned into $90,000$ samples for training,  $10,000$ each for validation and test. Each frame is composed of $256$ snapshots, which are used to compute the autocorrelation for a given frame. The test set is quarantined while the machine is trained on the training set and validated using the validation set repeatedly. Once the machine has achieved a saturation, the training is stopped and the test set is used to evaluate the model and generate the evaluation statistics. The number of sources to resolve is varied between $0$ or no source present in the sampled waveform to $9$ or the maximum detection capability for a ten element array. The classification accuracy on the test set was found out to be $89\%$.
\begin{table}[t!]
\centering
\begin{tabular}{cSSS} 
\toprule
& \multicolumn{3}{c}{CNNDetector}  \\
\cmidrule(r){2-4}
 {Class} & {Precision} & {Recall} & {F1} \\
\midrule
Class 0 & 1 & 1 & 1\\
Class 1 & 1 & 1 & 1\\
Class 2 & 0.99 & 1 & 0.99\\
Class 3 & 0.97 & 0.99 & 0.98\\
Class 4 & 0.93 & 0.95 & 0.94\\
Class 5 & 0.89 & 0.91 & 0.90\\
Class 6 & 0.86 & 0.86 & 0.86\\
Class 7 & 0.79 & 0.79 & 0.79\\
Class 8 & 0.75 & 0.69 & 0.72\\
Class 9 & 0.80 & 0.82 & 0.81\\
\bottomrule
\end{tabular}
\vspace{4mm}
\caption{Precision, recall and f-1 score obtained for experiment with non coherent sources for Detection framework-I.}
\label{table:acc_metrics}
\end{table}

 However, classification accuracy is not the ideal metric to evaluate a multi-class classification problem \cite{Goodfellow, murphy}. To evaluate further, we compute and tabulate the precision, recall and f-1 score in Table \ref{table:acc_metrics}. Precision is formerly defined as the positive predicted value and recall is the percentage of true positives that were correctly classified. The f-1 score is the harmonic mean between these two values \cite{p_r_f1}.      

\begin{figure}[H]
\center
\includegraphics[scale=0.7]{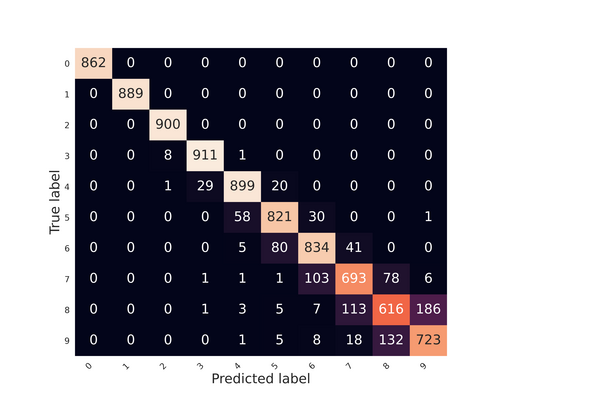}
\caption{Confusion matrix for CNNDetector. Rows in the table correspond to the True labels and columns are the predicted labels.}
\label{fig:Cmap}
\end{figure}

Finally, the performance of the framework on the test set is visualized using a confusion matrix in Figure \ref{fig:Cmap}. Each row in this matrix is the actual number of occurrence of each class in our test set, and each column is the respective classification by the algorithm \cite{hastie01}. In the confusion matrix we observe that the network accuracy of classification decreases as the number of sources increases. Samples with low number of sources (0-2) had none of the data miss-classified. In agreement with our precision and recall values, the classification error increases with an increasing number of sources to resolve. For a large presence such as with $8$ sources, CNNDetector correctly classifies only $616$ out of $931$ $(66.16$\%$)$ instances of the specific class in the test data. It fares slightly better with the class  of 9 sources, where it improves its performance and classifies $723$ out of $887$ $(81.5$\%$)$ samples correctly. This can be explained by the reasoning that the extreme tail end of the distribution (maximum number of sources) is relatively more easy for mapping and thus the machine error is less as compared to 7 or 8 sources.  This reasoning is also validated by the precision, recall and f-1 scores for each class. To the best of the authors knowledge, no other literature has reported the performance of detection for such high number of sources present in the sampled waveform \cite{feifeinet,cnn_sourcedetect}, hence it is not possible to draw a reference.
\begin{figure}[b!]
\center
\includegraphics[scale = 0.33]{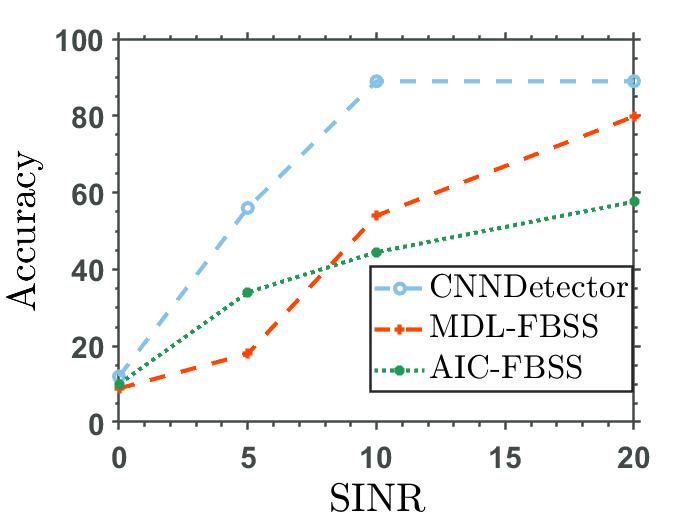}
\caption{Accuracy as a function of different SINR for the experiment involving 0-9 sources.}
\label{fig:sinr}
\end{figure}
\begin{figure}[H]
\center
\includegraphics[scale = 0.33]{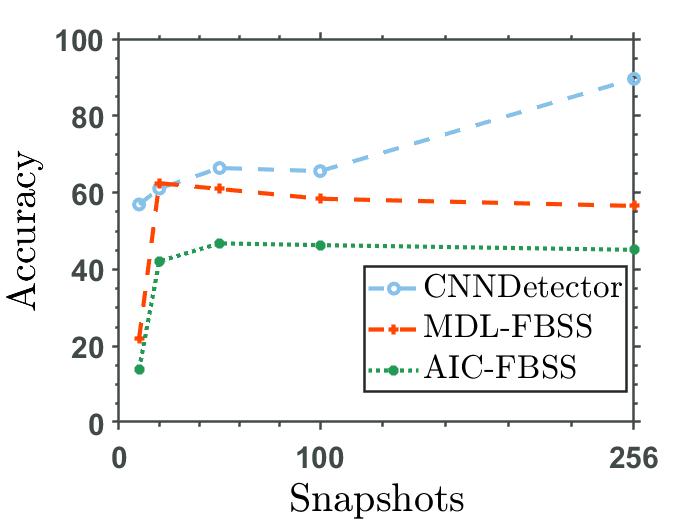}
\caption{Accuracy versus number of snapshots for 0-9 non correlated sources.}
\label{fig:snap}
\end{figure}

Next, we evaluate the CNNDetector from an operational point of view by varying SINR conditions. For any realistic scenarios, the SINR will vary and this does have a huge effect on any prediction model. We vary the SINR between $0dB$ to $20dB$ in four incremental steps and compare the detection performance to AIC and MDL. The corresponding accuracy for each of the discrete SINR thresholds probed in this research is presented in Figure \ref{fig:sinr}. We observe that for $0$ dB SINR, the accuracy of the CNNDetector is comparable to the conventional methods of MDL and AIC. But even with the slightest increase in signal power over the noise floor, the CNNDetector performs much better in term of accuracy than the conventional methods for a wide range of SINR until MDL catches up with CNNDetector when the SINR reaches $20$ dB. The accuracy shown in Figure \ref{fig:sinr} is obtained by averaging the accuracy over all classes. 

The frame size was fixed at $256$ for the results presented above. However, we can perform detection with fewer number of snapshots in a frame, but the threshold is ambiguous at this point. To answer this research question we train different models by varying the frame size. We sweep the frame size between $10$ and $256$ for this experiment to analyze the change in performance for varying frame sizes. The corresponding accuracy for each of the discrete snapshot sizes investigated in this research is presented in Figure \ref{fig:snap}. The results are both surprising and intuitive. It turns out that the conventional methods such as MDL and AIC saturates at a relatively smaller number of snapshots around $20$. The CNNDetector on the other hand is able to leverage the increasing number of snapshots available to it and keeps improving until we each the max threshold of $256$. There is no reason to believe that this behavior will not continue if we keep increasing the framesize.

\section{Detection Framework II}
\subsection{RadioNet}
The basic building block of  RadioNet is a ResNet module introduced in \cite{resnet}. Resnet models perform better than traditional CNNs because it is easier to optimize the residual mapping than to optimize the original complex desired mapping. The residual blocks form the backbone of the ResNet architecture and has shown to outperform conventional CNNs for a host of machine learning and computer vision applications \cite{resnet,resnet_vs_vgg}. Increasing the number of layers result in very deep conventional networks which causes a saturation in learning at a threshold and the network with even greater depth becomes very difficult to optimize. ResNets solve this issue to a great extent by introducing residual blocks. Instead of learning the unreferenced function the network has an easier time learning the residual function with reference to the layer inputs. The output of the residual block can be written as,
\begin{equation}
    {\bf{y}} = {\mathcal{F}}(\bf{x}) + x
\end{equation}
where, ${\bf{x}}$ is the input to the block, and ${\bf{y}}$ the output. The block is trained to learn $\mathcal{F}({\bf{x}})$, which is less complex than the desired output ${\mathcal{F}}(\bf{x}) + x$. Then at the output, an identity mapping is performed and the input is added to $\mathcal{F}({\bf{x}})$. This forms the residual block. Stacking multiple such residual blocks were shown to decrease the training error faster while being easier to optimize as compared to a traditional CNN of same depth \cite{resnet}. 

\begin{figure}
    \includegraphics[width=\linewidth]{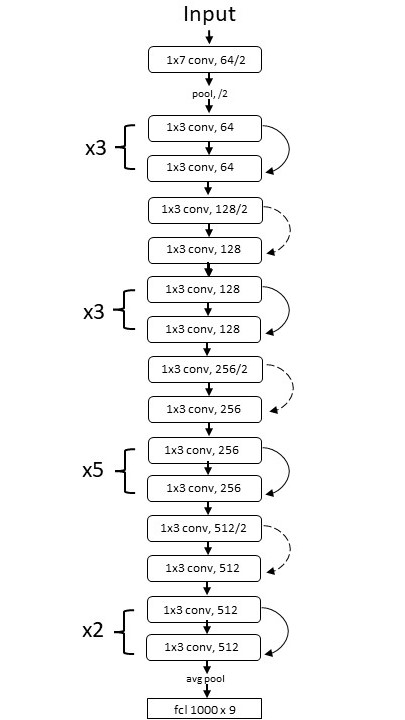}
    \caption{Network architecture of RadioNet. The dotted arrows increase dimensions.}\label{fig:radionet_arch}
  \end{figure}

RadioNet is a ResNet34 model which consists of $16$ residual blocks performing identity mapping with average pooling done before culminating into fully connected layers at the end. Small changes are made to the traditional Resnet34 architecture to accommodate our one-dimensional input data. The size of the filter is $[1 \times 7]$ in the first layer, and then kept constant as $[1 \times 3]$. The number of filters vary from 64,128,256 and 512. These hyper-parameters are optimized using a grid search. The final architecture is as shown in Figure \ref{fig:radionet_arch}. We use the same categorical cross entropy loss function to train our network. The computed errors are backpropagated during training phase and used to update the weights with a learning rate of $0.0001$. Stochastic Gradient Descent (SGD) is used to optimize the network parameters with a Nesterov momentum of value $0.9$ in batches of size $64$ \cite{SGD}. Once the network is trained, it is tested for correlated sources as shown in Algorithm \ref{algo:Source Detection in the presence of correlated signals using RadioNet}.    
\begin{algorithm}
  \begin{algorithmic}[1]\label{algo:Source Detection in the presence of correlated signals using RadioNet}
   \STATE \textbf{Require: }{$\bf{X}$, $L$, $RadioNet$, $K$} \\
   \STATE \textbf{Outputs: }{$y_{label}$}\\
   \STATE $temp$ $\in$$\mathbb{C}^{1 \times (L.(L+1/2))}$ $\gets$ random temporary array
    \STATE j=K
    \FOR{$i \in \{1,\dots,k\}$} 
     \STATE $\bf{X_f} = \bf{X[i:j,:]} $\\
     \STATE $\bf{X_b} = \bf{X[L-j+1:L-j+k,:]}$\\
     \STATE $\bf{R_f[i]} = \bf{X_f.X_f^H}$\\
     \STATE $\bf{R_b[i]} = \bf{X_b.X_b^H}$\\
     \STATE $j += 1$
     \ENDFOR
   
   \FOR{$i \in \{1,\dots,k $\}}
    \STATE $\bf{R_{fb}}$ = $\bf{R_f[i]}$ + $\bf{R_b[i]}$ \\
   \ENDFOR
   \STATE $ \bf{R_{fb}} = \bf{R_{fb}}/(2 \times K)$
   \STATE l,m = 1
   
    \FOR{$i$ $\in$ \{$1$,\dots,$L$\}}
      \FOR{$j \in \{l,\dots,L$\}}
        \STATE temp[m]  = $\bf{R_{fb}}$[i,j] \\
        \STATE m += 1
      \ENDFOR
      \STATE l += 1
    \ENDFOR\\
       
  \STATE j = 1 \\
  \FOR{$j \in \{1,\dots,(L\times(L+1)/2)$\}}
    \STATE feature vector[i] = $\mathbb{R}(temp[j])$\\
    \STATE i += 1\\
    \STATE feature vector[i] = $\mathbb{I}(temp[j])$\\
    \STATE i += 1\\
    
   \ENDFOR
   \STATE $y_{label}$ $\xleftarrow{}$ $RadioNet$(feature vector)\\
   \RETURN $y_{label}$
  \end{algorithmic}
  \caption{Source number detection in the presence of coherent sources using RadioNet}
\end{algorithm}

\subsection{Experiments and Evaluation}

For the second set of experiments, the dataset consisting of correlated sources was considered. In each frame, the number of non coherent sources is kept varying from 0 to 5 and the number of correlated sources is varied from 0 to 4. To perform FBSS on the sampled signals, the number of elements considered in the subarray is 5. The network is first trained on the first dataset for 10 epochs, and then later fed the second dataset consisting of correlated sources. In this experiment the CNNDetector was found to have an accuracy of only $56.7\%$. This was improved by RadioNet which obtained a validation accuracy of $82\%$ after 100 epochs. The precision, recall and f-1 scores were also calculated as shown in Table \ref{table:acc_metrics2}. We can see from this table that CNNdetector performs poorly in the presence of correlated sources, but RadioNet is still able to resolve upto 5 sources in the presence of correlated sources with a precision of $0.7$. 

\begin{table}
\centering
\begin{tabular}{cSSSSSS} 
\toprule
& \multicolumn{3}{c}{CNNDetector}  & \multicolumn{3}{c}{RadioNet} \\
\cmidrule(r){2-4}\cmidrule(l){5-7}
 {Class} & {Precision} & {Recall} & {F1} & {Precision} & {Recall} & {F1} \\
\midrule
Class 0 & 1 & 1 & 1 & 1 & 1 & 1\\
Class 1 & 0.94 & 1 & 0.97 & 0.99 & 1 & 0.99\\
Class 2 & 0.60 & 0.86 & 0.71 & 0.99 & 1 & 0.99\\
Class 3 & 0.42 & 0.46 & 0.71 & 0.97 & 0.99 & 0.98\\
Class 4 & 0.36 & 0.07 & 0.12 & 0.63 & 0.65 & 0.64\\
Class 5 & 0.46 & 0.55 & 0.5 & 0.7 & 0.58 & 0.64\\
\bottomrule
\end{tabular}
\vspace{4mm}
\caption{Precision, recall and f-1 score obtained for experiment with correlated coherent sources for CNNDetector and RadioNet.}
\label{table:acc_metrics2}
\end{table}
\begin{figure}
\center
\includegraphics[scale = 0.33]{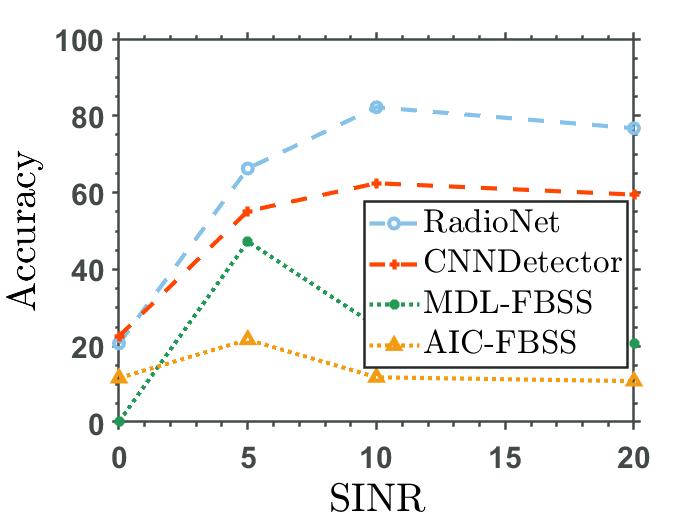}
\caption{Accuracy as a function of different SINR for the experiment involving 0-5 coherent sources.}
\label{fig:sinr_correlated}
\end{figure}

We also study the effect SINR has on the model performance. Similar to the previous experiment, we run both our networks along with FBSS-MDL and FBSS-AIC for data collected of differing SINRs to compare the accuracy, as shown in Figure \ref{fig:sinr_correlated}. We observe that both our deep learning based models consistently perform better than both MDL and AIC with FBSS. At higher SINR values RadioNet seems to perform far more better than MDL, AIC and CNNDetector, and hence seems to be the best choice for source detection.  
\begin{figure*}
\center
\includegraphics[scale=0.2]{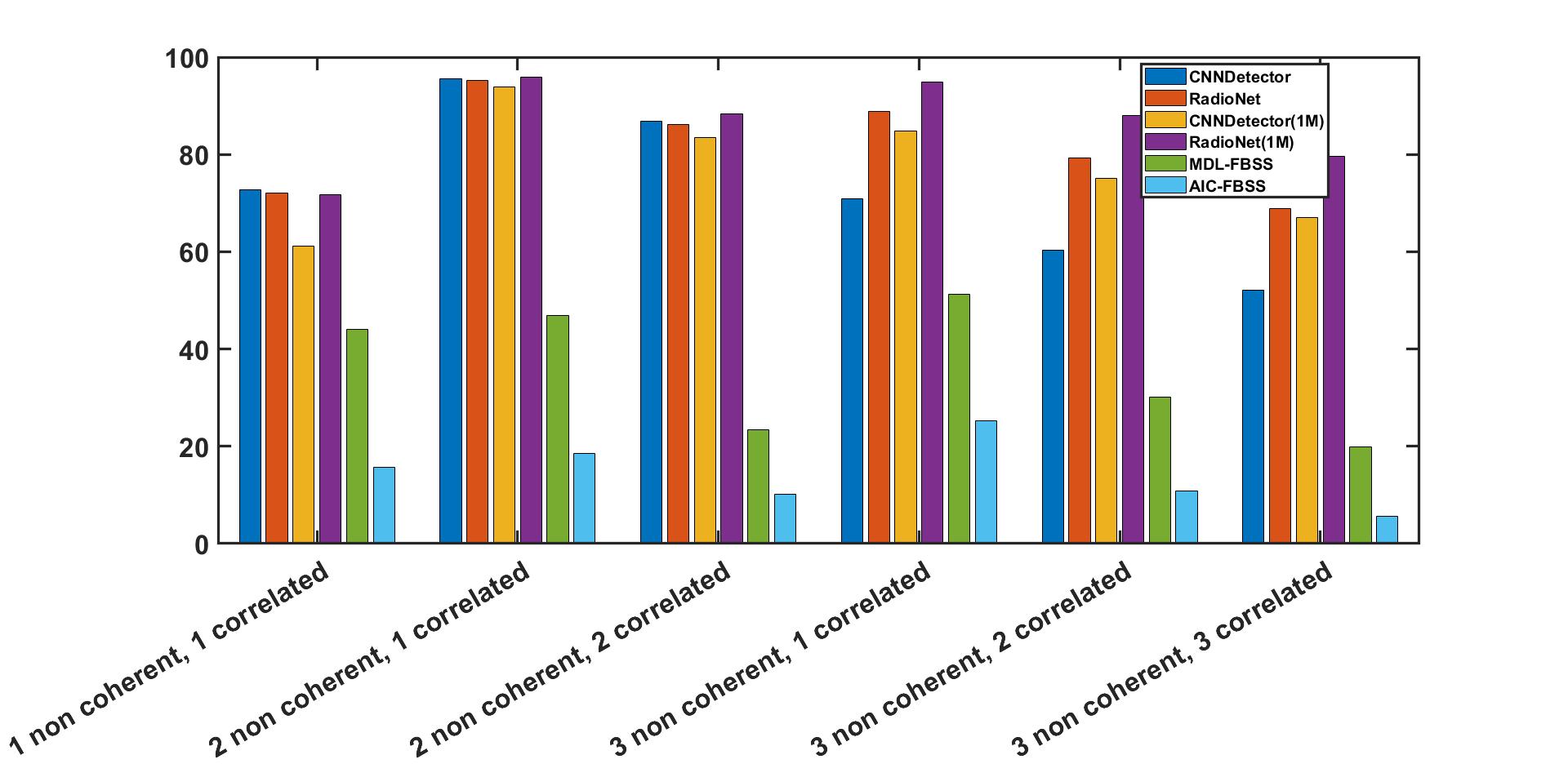}
\center
\caption{Accuracy of CNNDetector and RadioNet as compared with MDL and AIC for varying number of correlated sources.}
\label{fig:bar}
\end{figure*}

\begin{table*}
\begin{center}
\begin{tabular}{ccSSSSS} 
\toprule
{Number of sources} & \multicolumn{2}{c}{CNNDetector} & \multicolumn{4}{c}{RadioNet} \\
\cmidrule(r){2-4}\cmidrule(l){5-7}
{}  & {Precision} & {Recall} & {F1} & {Precision} & {Recall} & {F1} \\
\midrule
0  & 1 & 1 & 1 & 1 & 1 & 1 \\
1  & 0.99 & 1 & 0.99 & 0.99 & 0.99 & 0.99\\
2  & 0.85 & 0.90 & 0.88 & 0.90 & 0.94 & 0.92\\
3  & 0.75 & 0.79 & 0.77 & 0.83 & 0.84 & 0.84 \\
4  & 0.70 & 0.68 & 0.69 & 0.79 & 0.72 & 0.75 \\
5  & 0.74 & 0.68 & 0.71 & 0.77 & 0.79 & 0.78 \\
\bottomrule
\end{tabular}
\vspace{4mm}
\caption{Precision, recall and f-1 score obtained for 0-5 correlated sources with CNNDetector and RadioNet trained on one million samples.}
\label{table:correlated}
\end{center}
\end{table*}
Then we proceed to study the effect of the number of correlated sources affects the model performance in classifying each source. We create different datasets of varying number of coherent and correlated signals consisting of 3000 samples and test the trained models on it. We also use this dataset and test it on FBSS-MDL and FBSS-AIC. All these results obtained for each class are summarized in Figure \ref{fig:bar}. As seen in this plot, CNNDetector is the most suitable model while considering 1 or 2 non coherent signals. However its performance is surpassed by RadioNet when we increase the number of signals to 3. This can be understood by realizing that simpler networks are sufficient to solve problems of low complexity. The strength of resnet models is apparent only when we realize complex problems which traditional CNN models cannot solve. Both the deep learning solutions we propose have better accuracy than conventional techniques like MDL and AIC. We also see that the performance of our models tends to deteriorate if the number of correlated signals gets closer to the number of non coherent signals. Hence, it is shown that the proposed deep learning based models can perform source detection only when their number is less than the non-coherent signals. Finally the experiment was repeated with 1 million data points used for training. We also calculate the precision, recall and f-1 scores for each source in this experiment, as shown in Table \ref{table:correlated}.  This ten fold increase in the training dataset resulted in negligible changes in the CNNDetector performance. However RadioNet accuracy jumps up for each class. It attains an accuracy of $88\%$ for 3 non coherent, 2 correlated case. Despite this improvement it only obtains an accuracy of $79.9\%$ for 3 non coherent, 3 correlated case. Hence even RadioNet trained on one million samples is unsuccessful to resolve the number of sources when the number of coherent sources is equal to the number of correlated sources.

\section{Conclusion}
We introduced two modified CNN architectures referred to as the CNNDetector and RadioNet in this paper. The networks are trained to detect the number of sources sampled by an antenna array for a given time instance. We conduct extensive evaluation of the network with a large and diverse test set which can accurately identify for up to 9 sources for an array of 10 elements operating at a SINR as low as 5 dB. In a second set of experiments, RadioNet was shown to perform source detection even in the presence of correlated sources. The network performance  was evaluated and RadioNet was shown to be better than all existing statistical and machine learning techniques. 

\section{Reproducibility}
The  datasets,  code  and  results  can  be  found  at  the Github repository  dedicated  to  this  work:  https://github.com/jkrishnan95v/Signal\_detector

\section*{Acknowledgment}
The authors would like to thank The University of New Mexico Center  for  Advanced  Research  Computing,  supported  in  part by  the  National  Science  Foundation,  for  providing  the  high-performance computing resources used in this work.

\ifCLASSOPTIONcaptionsoff
  \newpage
\fi

\bibliographystyle{IEEEtran}
\bibliography{SourceDetection}
\end{document}